\documentclass{aa}  

\usepackage{graphicx}
\usepackage{txfonts}
\usepackage{hyperref}
\def\hseventy{h_{70}}
\def\dtobs{\Delta t_{2,1}^{(\mathrm{obs})}}
\def\ddt{D_{\Delta t}}
\def\mobs{M_*^{\mathrm{(obs)}}}

\def\mstar{M_*}
\def\mstartrue{M_*^{\mathrm{(true)}}}
\def\mstartrueein{M_{*,\mathrm{Ein}}^{\mathrm{(true)}}}

\def\detJ{\mathrm{det}J}

\def\msps{M_*^{\mathrm{(sps)}}}

\def\mfive{M_{\mathrm{DM},5}}

\def\tein{\theta_{\mathrm{Ein}}}

\def\asymm{\xi_{\mathrm{asymm}}}
\def\rmur{r_{\mu_r}}
\def\rmurobs{r_{\mu_r}^{(\mathrm{obs})}}
\def\gammadm{\gamma_{\mathrm{DM}}}

\def\asps{\alpha_{\mathrm{sps}}}

\def\betamax{\beta_{\mathrm{max}}}
\def\betasl{\beta_{\mathrm{SL}}}
\def\betaein{\beta_{\mathrm{Ein}}}

\def\mhalo{M_{200}}

\def\reff{R_{\mathrm{e}}}

\def\hyperpars{\boldsymbol{\eta}}
\def\indpar{\boldsymbol{\psi}}
\def\indpari{\boldsymbol{\psi}_i}

\def\data{\mathbf{d}}
\def\datai{\mathbf{d}_i}

\def\Sref#1{Section~\ref{#1}\xspace}
\def\Fref#1{Figure~\ref{#1}\xspace}
\def\Tref#1{Table~\ref{#1}\xspace}
\def\Eref#1{Equation~\ref{#1}\xspace}

\def\pr{{\rm P}}

\setcitestyle{notesep={}}

\defcitealias{S+C21}{Paper~I}

\begin{document}

   \title{Statistical strong lensing. II. Cosmology and galaxy structure with time-delay lenses}
   \titlerunning{Statistical strong lensing. II.}
   \authorrunning{Sonnenfeld}


   \author{Alessandro Sonnenfeld\inst{1}
          }

   \institute{Leiden Observatory, Leiden University, Niels Bohrweg 2, 2333 CA Leiden, the Netherlands\\
              \email{sonnenfeld@strw.leidenuniv.nl}
             }

   \date{}

 
  \abstract
    {
Time-delay lensing is a powerful tool for measuring the Hubble constant $H_0$. However, in order to obtain an accurate estimate of $H_0$ from a sample of time-delay lenses, very good knowledge of the mass structure of the lens galaxies is needed. Strong lensing data on their own are not sufficient to break the degeneracy between $H_0$ and the lens model parameters on a single object basis.
}
   {
The goal of this study is to determine whether it is possible to break the $H_0$-lens structure degeneracy with the statistical combination of a large sample of time-delay lenses, relying purely on strong lensing data with no stellar kinematics information.
} 
   {
I simulated a set of 100 lenses with doubly imaged quasars and related time-delay measurements. I fitted these data with a Bayesian hierarchical method and a flexible model for the lens population, emulating the lens modelling step.
}
   {
The sample of 100 lenses on its own provides a measurement of $H_0$ with $3\%$ precision, but with a $-4\%$ bias.
However, the addition of prior information on the lens structural parameters from a large sample of lenses with no time delays, such as that considered in Paper I, allows for a $1\%$ level inference.
Moreover, the 100 lenses allow for a $0.03$~dex calibration of galaxy stellar masses, regardless of the level of prior knowledge of the Hubble constant.
}
   {
Breaking the $H_0$-lens model degeneracy with lensing data alone is possible, but $1\%$ measurements of $H_0$ require either many more than 100 time-delay lenses or knowledge of the structural parameter distribution of the lens population from a separate sample of lenses.
}

   \maketitle
%

\section{Introduction}

Gravitational lensing time delays, the difference in light travel time between two or more strongly lensed images of the same source, are powerful probes of cosmology \citep{T+M16}.
For a single strong gravitational lens, the time delay between the multiple images of the background source depends on the lens mass distribution and on the geometry of the lens--source system, which in turn depends on the structure and expansion history of the Universe. In particular, in a Universe described by a Friedman-Lema\^{i}tre-Robertson-Walker metric, time delays scale with the inverse of the Hubble constant $H_0$ \citep{Ref64}.
The typical time delays of galaxy-scale strong lenses are on the order of weeks \citep[see e.g.][]{Mil++20b}, and can be measured if the flux from the background source varies in time.
However, knowledge of the mass distribution of the lens is required in order to convert a time-delay measurement into an estimate on $H_0$.

Some lens properties, such as the total projected mass enclosed within the multiple images, can be inferred very accurately by modelling the strongly lensed images. However, there is a fundamental limit to how much information can be extracted from strong lensing data alone, which is set by the mass-sheet degeneracy \citep{FGS85}: given a model that reproduces the image positions and magnification ratios of a strongly lensed source, it is always possible to define a family of alternative models that leave those observables unchanged and predict different time delays.

There are a few ways to break the degeneracy between the lens mass profile and $H_0$. One possibility is to assert a lens model that artificially breaks the mass-sheet degeneracy, such as, for example, a model with a density profile described by a single power law. 
The parameters of a power-law lens model can be unambiguously constrained with lensing data alone, provided that the images of the background source can be well resolved \citep[see e.g.][]{Suy12}. 
By modelling a time-delay lens with a power-law lens model, it is possible to estimate $H_0$ directly given the observed time delays. However, if the true density profile of the lens is different from a power law, that estimate will be biased \citep[see e.g.][]{S+S13}.

Another possibility is to use stellar kinematics data to further constrain the mass model of the lens. This is the approach used in the latest analysis of the TDCOSMO collaboration\footnote{\url{http://tdcosmo.org}} by \citet{Bir++20}. In their study of seven time-delay lenses, \citet{Bir++20} combined lensing data with stellar kinematics observations consisting for the most part of measurements of the central velocity dispersion of the lens galaxy, and obtained a measurement of $H_0$ with 8\% precision. The addition of prior information from similar measurements on gravitational lenses with no time delays allowed these latter authors to reduce the uncertainty to 5\%, and the precision can be further improved by replacing central velocity dispersion measurements with spatially resolved kinematics data \citep{YSH20,B+T21}.

Nevertheless, in order to incorporate stellar kinematics constraints into a time-delay lensing study, it is necessary to model a variety of additional aspects of the lens system: most critically, the full 3D mass distribution of the lens galaxy and the phase-space distribution of the kinematical tracers (i.e. the stars that contribute to the observed spectrum). This can be very challenging, especially if percent precision on the measurement of $H_0$ is required.

In this paper, I explore an alternative method for breaking the degeneracy between the lens mass profile and $H_0$ based on the statistical combination of a large number of time-delay lenses.
Building on the work of \citet[][, hereafter Paper I]{S+C21}, I use a Bayesian hierarchical inference approach to simultaneously fit strong lensing data from a large sample of lenses including time-delay observations.
As in \citetalias{S+C21}, the challenge is to find a model that is sufficiently flexible to allow for an accurate inference of the key parameters, most importantly $H_0$, while not be so flexible that it can no longer be constrained without the need for stellar kinematics data.

While there are currently only a few lenses with the necessary data to carry out a time-delay analysis, the number of known strongly lensed quasars that could be followed-up and used for this purpose is already greater than 200\footnote{\url{https://research.ast.cam.ac.uk/lensedquasars/}}, and is expected to grow steadily thanks to new surveys like Euclid\footnote{\url{https://www.euclid-ec.org/}} and the Large Synoptic Survey Telescope (LSST\footnote{\url{https://www.lsst.org/}}). Most importantly, the LSST will enable the measurement of hundreds of time delays \citep{Lia++15} in virtue of its observations over hundreds of epochs with a cadence of a few days.
This paper lays out a strategy for the optimal use of these data.

Here I test this approach on simulated data for a set of 100 time-delay lenses. As in \citetalias{S+C21}, I emulate the lens-modelling process: instead of simulating and modelling lens images in full detail ---a lengthy process in terms of both human and computational time---, I compress the information that can be obtained via modelling with a handful of summary observables.
With this choice it is possible to focus on the statistical aspect of the problem, leaving aside the technical challenges associated with lens modelling.
The recent work of \citet{Par++21} on Bayesian neural networks offers a viable solution to such challenges.

Here, I address three questions: the extent to which we can constrain $H_0$ with strong lensing information from a sample of 100 time-delay lenses on its own; how the inference improves with the addition of prior information from a larger set of strong lenses with no time-delay measurements, such as the sample considered in \citetalias{S+C21};
and finally, what can be learned about the structure
of massive galaxies by combining external information ---the value of $H_0$
 known from a different experiment--- with a sample
of 100 time-delay lenses.

The structure of this paper is as follows. In \Sref{sect:theory} I explain the basics of lensing time delays. In \Sref{sect:sims} I present the simulations. In \Sref{sect:model} I describe the model that I fit to the simulated data. In \Sref{sect:results} I show the results of the experiment. I discuss the findings and draw conclusions in \Sref{sect:discuss}.

A flat Lambda cold dark matter cosmology with matter energy density $\Omega_M=0.3$ and cosmological constant $\Omega_\Lambda = 0.7$  is assumed throughout the paper. With this choice, I reduce the degrees of freedom in the cosmological model to the value of the Hubble constant alone.
For the creation of the simulated data, I assume $H_0=70$~km~s$^{-1}$~Mpc$^{-1}$.
The Python code used for the simulation and analysis of the lens sample can be found in a dedicated section of a GitHub repository\footnote{\url{https://github.com/astrosonnen/strong_lensing_tools}}.


\section{Lensing time delays}\label{sect:theory}

In this section, I present the theoretical foundation for the simulation and modelling of time-delay data. For an introduction to the strong lensing formalism, including an overview of the image configurations produced by axisymmetric lenses, I refer to Section 2 of \citetalias{S+C21}.

Let us consider a point source at angular position $\boldsymbol\beta$, gravitationally lensed by a single lens plane with lensing potential $\psi(\boldsymbol\theta)$. 
If $\boldsymbol\theta_1$ is the position of one of the images associated to the source, then the difference in the light travel time with respect to the case without lensing is
\begin{equation}\label{eq:timetravel}
t(\boldsymbol\theta_1) = \frac{\ddt}{c}\left[\frac{(\boldsymbol\theta_1 - \boldsymbol\beta)^2}{2} - \psi(\boldsymbol\theta_1)\right].
\end{equation}
In the above equation, $c$ is the speed of light and $\ddt$ is the time-delay distance, defined as
\begin{equation}
\ddt \equiv (1+z_d)\frac{D_\mathrm{d} D_\mathrm{s}}{D_\mathrm{ds}},
\end{equation}
where $z_d$ is the lens redshift, and $D_{\mathrm{d}}$, $D_{\mathrm{s}}$, and $D_{\mathrm{ds}}$ are the angular diameter distances between observer and lens, observer and source, and lens and source, respectively.
In a Friedman-Lema\^{i}tre-Robertson-Walker Universe, $\ddt$ scales with the inverse of the Hubble constant $H_0$.

The part in square brackets in \Eref{eq:timetravel} is the sum of two terms: the first is a geometrical term describing the increase in the light travel time due to the extra distance covered by a light ray compared to a straight line; the second term is a delay due to the lens potential.
When no lensing is present, $\boldsymbol\theta = \boldsymbol\beta$ and both terms go to zero.
From \Eref{eq:timetravel}, it follows that if the source is strongly lensed into an additional image at angular position $\boldsymbol\theta_2$, then the time delay $\Delta t_{2,1}$ between the two images is
\begin{equation}\label{eq:timedelay}
\Delta t_{2,1} = \frac{\ddt}{c}\left[\frac{(\boldsymbol\theta_2 - \boldsymbol\beta)^2}{2} - \psi(\boldsymbol\theta_2) - \frac{(\boldsymbol\theta_1 - \boldsymbol\beta)^2}{2} + \psi(\boldsymbol\theta_1)\right].
\end{equation}

In summary, the time delay between two images is the product of a term that depends on cosmology, $\ddt/c$, and a part that depends on the lens configuration and mass distribution, the term in square brackets. Of the quantities that enter the latter term, only the image positions $(\boldsymbol\theta_1,\boldsymbol\theta_2)$ are directly observable: the lens potential $\psi$ and the source position $\boldsymbol\beta$ must be inferred via lens modelling.

\subsection{Mass-sheet transformations}\label{ssec:masssheet}

Given a lens model ---consisting of a lens potential $\psi(\boldsymbol\theta)$ and a source position $\boldsymbol\beta$--- that reproduces all of the observed image positions and magnification ratios between images, the following class of transformations
\begin{align}\label{eq:mst}
\psi(\boldsymbol\theta) \rightarrow \lambda \psi(\boldsymbol\theta) + \frac{(1-\lambda)}{2}|\boldsymbol\theta|^2, \\
\beta \rightarrow \lambda\beta,
\end{align}
leaves those observables unchanged.
 \Eref{eq:mst} is called a mass-sheet transformation. 
The fact that image positions and magnification ratios alone can only constrain a lens model up to a transformation of this kind is referred to as the mass-sheet degeneracy.
 
Time delays on the other hand are not invariant under a mass-sheet transformation. 
By applying the transformation of \Eref{eq:mst} to \Eref{eq:timedelay}, I find that the time delay between the two images transforms as
\begin{equation}
\Delta t_{2,1} \rightarrow \lambda \Delta t_{2,1}.
\end{equation}
This means that, without any assumptions on the lens mass distribution, it is not possible to use image positions and magnification ratios to unambiguously predict the time delay between two images.
The strategy that I propose to break the mass-sheet degeneracy consists in the adoption of physically motivated lens models and on the statistical combination of a large number of lenses.



\section{Simulations}\label{sect:sims}

The experiment is carried out on a sample of 100 simulated lenses. I generated this sample using a very similar prescription to that of \citetalias{S+C21}, then added time-delay measurements.
In this section, I summarise the procedure used to create this simulation and show the time-delay distribution of the sample.

\subsection{Properties of the lens systems}\label{ssec:simprop}

All of the lenses are assumed to be isolated and to have an axisymmetric mass distribution.
Moreover, for the sake of reducing the computational burden of the analysis, all of the lenses are at the same redshift $z_d=0.4$ and all of the sources are at $z_s=1.5$.
Each lens consists of the sum of a stellar component ---described by a de Vaucouleurs profile--- and a dark matter halo. The density profile of the dark matter halo is determined following the prescription of \citet{Cautun2020}: starting from a Navarro, Frenk \& White \citep[NFW][]{NFW97} halo, the mass distribution is modified to simulate the contraction of the dark-matter distribution in response to the infall of baryons. The resulting dark-matter profile can be well approximated by a generalised NFW (gNFW) model with an inner density slope steeper than that of an NFW profile \citepalias[see Figure~3 in][]{S+C21}.

The source is approximated as a point. This is intended to describe both a time-varying active galactic nucleus (AGN) component, used for the measurement of the time delay, and its host galaxy. The point source approximation for the host galaxy is done to reduce the amount of data that needs to be simulated and modelled. In real time-delay lens analyses, the full surface brightness distribution of the source is used to constrain important parameters of the lens model \citep[see e.g.][]{Suy++13,Din++21}.
While a point source does not allow for such a measurement, I still take into account the information provided by the source surface brightness distribution by emulating the lens modelling process. Section \ref{ssec:data} explains how this is done.

The properties of each lens system are fully determined by the following set of parameters.
\begin{itemize}
\item The total stellar mass of the galaxy, $\mstartrue$.
\item The `stellar population synthesis-based stellar mass', $\msps$. This is the stellar mass corresponding to a stellar population synthesis model that reproduces the observed luminosity and colours of the galaxy in the absence of photometric errors. The relation between $\msps$ and $\mstartrue$ is described by the stellar population synthesis mismatch parameter, $\asps=\mstartrue/\msps$, which is in general different from unity because of unknown systematic uncertainties associated with the stellar population synthesis model. 
\item The galaxy half-light radius, $\reff$. I assume that the stellar mass surface mass density follows the surface brightness profile of the galaxy. Therefore, $\reff$ is also the half-mass radius of the stellar component.
\item The virial mass of the dark matter halo, $\mhalo$.
\item The source position with respect to the lens centre, $\beta$.
\end{itemize}

As a consequence of the axisymmetric lens assumption, every lens produces either two or three images of the background source. Only the two brighter images are considered for the analysis, as the third image, if present, is usually highly de-magnified. 
Although five of the seven time-delay lenses studied in the latest analysis of the TDCOSMO collaboration are quadruple (quad) lenses \citep{Bir++20}, which are qualitatively different from the doubly imaged systems of this simulation, doubles are expected to dominate over quads in a survey like the LSST by a factor of about six \citep{O+M10}.
Therefore, a sample consisting entirely of doubles is a good first approximation of upcoming samples of time-delay lenses.

Let us consider a Cartesian coordinate frame centred on the lens, with the $\hat{\mathbf{x}}$ axis aligned with the source position so that $\boldsymbol{\beta} = \beta\hat{\mathbf{x}}$ with $\beta > 0$.
For a lens with Einstein radius $\tein$, the two images are located at $\boldsymbol\theta_1 = \theta_1\hat{\mathbf{x}}$ with $\theta_1 > \tein$, and $\boldsymbol\theta_2 = \theta_2\hat{\mathbf{x}}$ with $-\tein < \theta_2 < 0$.
In other words, image 1, which is referred to here as the main image, is outside of the tangential critical curve, while image 2, the counter-image, is inside of the critical curve on the opposite side with respect to the lens centre.

\subsection{Sample creation algorithm}\label{ssec:algorithm}

The sample was created as follows. 
\begin{enumerate}
\item Values of $\log{\msps}$ were drawn from a Gaussian distribution set to approximate the stellar mass distribution of known lens samples.
\item The true stellar masses were obtained by setting the stellar population synthesis mismatch parameter to $\log{\asps}=0.1$ for all galaxies.
\item Half-light radii were assigned using a power-law mass-size relation with log-Gaussian scatter.
\item Values of the halo mass were drawn using a power-law scaling relation with $\msps$ and log-Gaussian scatter.
\item The concentration parameter (the ratio between the virial radius and the scale radius) of the initial (pre-contraction) NFW profile of the dark matter halo was set to $5$.
\item The \citet{Cautun2020} prescription was applied to obtain the contracted dark-matter density profile.
\item The position of the source $\beta$ was drawn from a uniform distribution within a circle of radius $\betamax$, where $\betamax$ was set in such a way as to avoid very asymmetric image configurations. In particular,  the asymmetry parameter $\asymm$ was considered, which is defined as
\begin{equation}
\asymm \equiv \frac{\theta_1 + \theta_2}{\theta_1 - \theta_2}.
\end{equation}
This quantity is $0$ for a lens with $\theta_2 = -\theta_1$, corresponding to the case in which the source is aligned with the optical axis ($\beta=0$), and increases as the relative position of the two images with respect to the lens centre becomes more asymmetric. I set $\betamax$ to be the value of the source position corresponding to a maximum allowed asymmetry parameter $\asymm=0.5$. This criterion is meant to simulate an arbitrary selection effect associated with the detectability of lenses: in a real survey, strong lenses can only be identified as such if multiple images are detected. In a lens with a highly asymmetric configuration, the counter-image is close to the centre and has typically low magnification, and is  therefore more difficult to detect. The choice of the $\asymm$ threshold is arbitrary, but does not affect the results.
\citetalias{S+C21} used a slightly different criterion to assign the source position based on the magnification of the inner image. I opted for a different criterion for this experiment, one that is not based on any magnification information, which is difficult to obtain in practice.
\end{enumerate}

\subsection{Observational data}\label{ssec:data}

I take the two image positions, $(\theta_1, \theta_2)$, to be measured exactly.
In addition, I simulate measurements of the radial magnification ratio between the two images, $\rmur$. For a lensed extended source, such as the host galaxy of a strongly lensed AGN or supernova, it is possible to measure the radial magnification ratio between the main image and its counter-image by comparing the widths of the two arcs. This information is usually implicitly obtained by modelling the full surface brightness distribution of the source, and is a model-independent constraint on the radial profile of the lens mass distribution \citep{Son18, Sha++21}.
By directly providing an estimate of $\rmur$, I am emulating the lens-modelling process.

I set the uncertainty on $\rmur$ to be $0.05$ based on the constraining power of Hubble Space Telescope data on the density profile of typical lenses, as indicated by the work of \citet{Sha++21}.
I am thus assuming a scenario in which imaging data with sufficient depth and resolution to detect and resolve the host galaxy are available for each lens.

\begin{figure*}
\includegraphics[width=\textwidth]{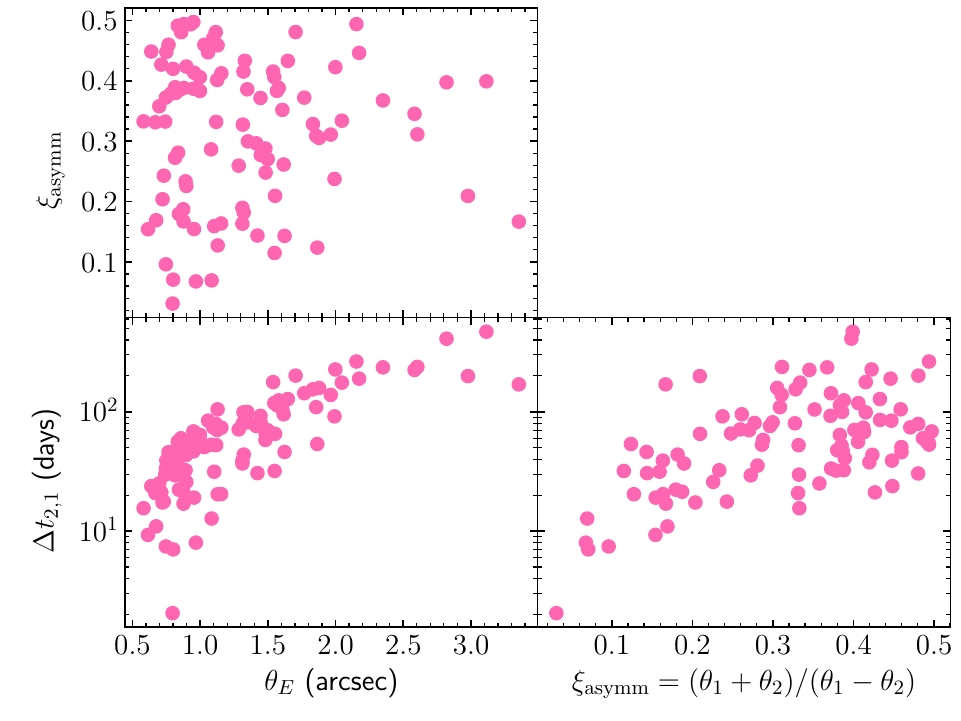}
\caption{
Distribution in the time delay between images 2 and 1, as well as the Einstein radius, and the image configuration asymmetry parameter $\asymm$ of the lens sample.
\label{fig:dtplot}
}
\end{figure*}

I then calculated the time delays of the sample.
The bottom-left panel of \Fref{fig:dtplot} shows the distribution of $\Delta t_{2,1}$ as a function of the Einstein radius. The two quantities are strongly correlated. This is expected given \Eref{eq:timedelay}: the time delay scales as the square of the angular scale of a lens, which in turn is set by the Einstein radius. 
The time delays of the sample span the range between $2$ and $469$~days, with a median value of $55$~days.

\Fref{fig:dtplot} also shows the distribution of the image configuration asymmetry parameter $\xi$. As the top-left panel of \Fref{fig:dtplot} shows, $\asymm$ does not correlate with the Einstein radius.
However, the time delay correlates with $\asymm$, as shown in the bottom-right panel of \Fref{fig:dtplot}: lenses with a more asymmetric image configuration tend to have longer time delays. In the limiting case of a perfectly symmetric image configuration ($\theta_2 = -\theta_1$), the time delay is zero.

For some of these lenses, it might be difficult to obtain precise time-delay measurements, in practice: for example, microlensing introduces a systematic error on the order of a day \citep{T+K18}, which limits the usefulness of the lenses with the smallest time delays.
For this reason, in a real time-delay lensing campaign one might wish to exclude lenses with a small Einstein radius and image configuration close to symmetric.\ I did not apply any such cut to the sample, for simplicity. 

I added a Gaussian observational noise to the time-delay measurements with a dispersion equal to $10\%$ of the true time delay. The uncertainty on a time delay of $50$~days is therefore $5$ days.
This value is similar to the uncertainties observed in the Time Delay Challenge II \citep{Lia++15}, which was designed on the basis of expectations for the LSST.
A constant relative error on the time delays is perhaps not very realistic, as it translates into an uncertainty of only a few hours for the lenses with the smallest values of $\Delta t_{2,1}$, which is difficult to achieve. 

However, this choice makes interpretation of the results easier (particularly the discussion of \Sref{sect:results}), as it ensures that each lens has equal weight on the inference of $H_0$. 
On the contrary, if I were to adopt a constant absolute uncertainty on $\Delta t_{2,1}$ in a sample such as this, which spans more than two orders of magnitude in the time delay, the lenses with the longest time delays would end up dominating the inference.
Nevertheless, I repeated the analysis assuming a five-day constant error on $\Delta t_{2,1}$ instead, finding very similar results to the ones shown in this paper.

I also assume the lens and source redshifts to be known exactly.
Finally, for each galaxy, I simulated a measurement of the stellar-population-synthesis-based stellar mass, $\msps$, with an uncertainty of $0.15$~dex.

In summary, for each lens I have the following data.
\begin{itemize}
\item The positions of the two brightest images of the source, $(\theta_1, \theta_2)$, measured exactly.
\item The lens and source redshift, known exactly.
\item The observed ratio of the radial magnification between images 1 and 2, $\rmurobs$, measured with an uncertainty of $0.05$.
\item The observed time delay between images 2 and 1, $\Delta t_{2,1}^{(\mathrm{obs})}$, measured with an uncertainty of $10\%$.
\item The observed stellar mass, $\mobs$, obtained from stellar population synthesis fitting and under the assumption of a reference value of the Hubble constant, measured with an uncertainty of $0.15$~dex.
\end{itemize}
Among these measurements, only the last one depends on assumptions on cosmology. In order to make it independent of $H_0$, I consider the following quantity instead:
\begin{equation}
h_{70}^2\mobs,
\end{equation}
where $h_{70}$ is the ratio between the assumed value of $H_0$ and the reference value of $70$~km~s$^{-1}$~Mpc$^{-1}$,
\begin{equation}
h_{70} \equiv \frac{H_0}{70\mathrm{km}\,\mathrm{s}^{-1}\,\mathrm{Mpc}^{-1}}.
\end{equation}
The fact that the reference value of $H_0$ is the same as the value used to generate the mock has no influence on the analysis.


\section{The model}\label{sect:model}

I want to fit a model to the mock data in order to simultaneously infer the distribution in the mass structure of the lenses and the value of $H_0$.
In \citet{Son18}, I showed that, in order to predict the time delay of a strong lens with an accuracy of $1\%$, a lens model with at least three degrees of freedom in the radial direction is needed.
Motivated by this result, I employ the same model used in \citetalias{S+C21}, which satisfies this requirement, with slight modifications to allow for the additional freedom in the value of the Hubble constant.
In this section, I describe the model, as well as the Bayesian hierarchical inference method that I use to fit it to the data.

\subsection{Individual lens model}\label{ssec:individ}

I describe each lens as the sum of a stellar component and a dark matter halo. I assume that the stellar profile of each lens can be measured exactly up to a constant scaling of the mass-to-light ratio. 
This is a reasonable assumption, as the light profile of a lens galaxy can be determined with high precision, at least in the region enclosed by the Einstein radius.

I describe the dark matter component with a gNFW profile, which is a model with q 3D density profile given by
\begin{equation}
\rho(r) \propto \frac{1}{(r/r_s)^{\gammadm}\left(1 + r/r_s\right)^{3-\gammadm}}.
\end{equation}
The parameter $\gammadm$ is the inner density slope. If $\gammadm=1$, the gNFW profile reduces to the NFW case.

A gNFW profile has three degrees of freedom: $\gammadm$, the scale radius $r_s$, and an overall normalisation. As in \citetalias{S+C21},  one degree of freedom is eliminated by fixing the value of the scale radius to
\begin{equation}
\hseventy r_s = 100\,\mathrm{kpc}.
\end{equation}
With this choice, the angular size of the scale radius is fixed and independent of the assumed value of $H_0$.
I parameterise the remaining two degrees of freedom of the dark matter component with the inner slope $\gammadm$ and the projected mass enclosed within an aperture of $5$~kpc~$\hseventy$, $\hseventy^2\mfive$.
Each lens system is then fully described by the following set of parameters: 
\begin{equation}
\indpar \equiv \left\{\frac{\mstartrue}{1/\hseventy^2}, \frac{\msps}{1/\hseventy^2}, \frac{\reff}{1/\hseventy}, \frac{\mfive}{1/\hseventy^2}, \gammadm, \beta \right\},
\end{equation}
where $\beta$ is the source position and $\reff$ is in physical units.

This model is underconstrained on an individual lens basis: the two image positions and the radial magnification ratio can only constrain three degrees of freedom, one of which must be the source position $\beta$. The time-delay measurement provides an additional constraint, but its interpretation depends on the value of $H_0$, which I want to infer.
As shown in Appendix~\ref{sect:appendixa}, the model does not break the mass-sheet degeneracy for physically allowed values of the mass-sheet transformation parameter $\lambda$.
The strategy here consists in statistically constraining only the overall properties of the lens population, together with $H_0$, rather than determining the exact structure of each lens.

\subsection{Population distribution}\label{ssec:popdist}

I assume that the parameters of each lens, $\indpar$, are drawn from a common probability distribution $\pr(\indpar|\hyperpars)$ describing the population, which is in turn summarised by a set of parameters $\hyperpars$. I choose the following functional form for the population distribution,
\begin{equation}\label{eq:popdist}
\begin{split}
\pr(\indpar|\hyperpars) = & \mathcal{S}\left(\frac{\msps}{1/\hseventy^2},\frac{\reff}{1/\hseventy}\right)\,\mathcal{A}\left(\frac{\mstartrue}{\msps}\right)\,\mathcal{H}\left(\frac{\mfive}{1/\hseventy^2}\right)\times \\ 
& \mathcal{G}(\gammadm)\,\mathcal{B}\left(\beta\left\vert\frac{\mstartrue}{1/\hseventy^2},\frac{\reff}{1/\hseventy},\frac{\mfive}{1/\hseventy^2},\gammadm\right.\right),
\end{split}
\end{equation}
which is the same used in \citetalias{S+C21}.

The first term, $\mathcal{S}$, describes the distribution in stellar mass and half-light radius of the lens sample. For the sake of simplifying calculations, I assume that it is known exactly. Therefore, I set it equal to the distribution used to generate the sample, which is a bi-variate Gaussian in $(\log{(\hseventy^2\msps)},\log{(\hseventy\reff)})$. This is a reasonable approximation, because this term can be constrained very well with a sample of 100 lenses \citep[see e.g.][]{Son++15}.

The term $\mathcal{A}$ describes the distribution in the stellar-population-synthesis mismatch parameter introduced in section \ref{ssec:simprop}, $\asps=\mstartrue/\msps$. I model this term as a Dirac delta function, that is, I assume a single value of $\asps$ for the whole lens sample:
\begin{equation}\label{eq:aterm}
\mathcal{A} = \delta\left(\frac{\mstartrue}{\msps} - \asps\right).
\end{equation}

The terms $\mathcal{H}$ and $\mathcal{G}$ describe the distribution in the two parameters related to the dark matter distribution, $\mfive$ and $\gammadm$.
I model the former as a Gaussian in $\log{(\hseventy^2\mfive)}$ and the latter as a Gaussian in $\gammadm$:
\begin{align}
\mathcal{H}\left(\frac{\mfive}{1/\hseventy^2}\right) & \sim \mathcal{N}_{\log{(\frac{\mfive}{1/\hseventy^2})}}\left(\mu_{\mathrm{DM}}\left(\frac{\msps}{1/\hseventy^2},\frac{\reff}{1/\hseventy}\right), \sigma_{\mathrm{DM}}^2\right)\label{eq:hterm} \\
\mathcal{G}\left(\gammadm\right) & \sim \mathcal{N}_{\gammadm}\left(\mu_\gamma\left(\frac{\msps}{1/\hseventy^2},\frac{\reff}{1/\hseventy}\right), \sigma_\gamma^2\right)\label{eq:gterm}.
\end{align}
One of the main results of \citetalias{S+C21} was the finding that, in order to obtain an accurate inference of the average properties of the dark matter distribution of a lens sample, it is necessary to allow for a dependence of the average dark matter mass and inner slope with all of the dynamically relevant properties of the lens galaxy: in this case, the stellar mass and half-light radius.
For this reason, I allow for the means of the Gaussian distributions in the above equations to scale with $\msps$ and $\reff$ as follows:
\begin{align}
\mu_{\mathrm{DM}} & = \mu_{\mathrm{DM},0} + \beta_{\mathrm{DM}}\left(\log{\left(\frac{\msps}{1/\hseventy^2}\right)} - 11.4\right) + \nonumber \\
& \xi_{\mathrm{DM}}\left(\log{\left(\frac{\reff}{1/\hseventy}\right)} - \mu_R\left(\frac{\msps}{1/\hseventy^2}\right)\right)\label{eq:muh} \\
\mu_\gamma & = \mu_{\gamma,0} + \beta_\gamma(\log{\frac{\msps}{1/\hseventy^2}} - 11.4) + \nonumber \\
& \xi_\gamma(\log{\frac{\reff}{1/\hseventy}} - \mu_R(\msps)),\label{eq:mug}
\end{align}
where $\mu_R$ is the average value of $\log{(\hseventy\reff)}$ for a lens with stellar mass $\hseventy^2\msps$.

The last term in \Eref{eq:popdist} is $\mathcal{B}$, which describes the distribution in the source position.
At fixed lens properties, this term determines the distribution in the asymmetry of the image configuration. This aspect is directly related to the selection function of the sample: when building the mock data, I imposed a condition on the maximum value of $\asymm$ to simulate a selection that disfavours highly asymmetric configurations with a de-magnified counter-image.
I assume an uninformative prior on the source position: given the lens mass model parameters, I assume that the source has equal prior probability of being anywhere within the source plane circle of radius $\betasl$, which is strongly lensed into multiple images. 
If the lens has a radial caustic, then this sets the value of $\betasl$. Otherwise, I set $\betasl$ to the value that produces a counter-image at a very small distance from the lens centre.
With this definition, the source prior term reads
\begin{equation}\label{eq:betadist}
\mathcal{B}\left(\beta\right) = \left\{\begin{array}{ll} \dfrac{2\beta}{\betasl^2} & \rm{if}\,0 < \beta < \betasl\\
& \\
0 & \rm{elsewhere}\end{array}\right. .
\end{equation}
The functional form of \Eref{eq:betadist} is the same as the probability distribution used to generate the sample, with one important difference: instead of truncating the distribution at the value $\betamax$ corresponding to $\asymm=0.5$, I use the more conservative value of $\betasl$.
This is a different choice compared to the analysis of \citetalias{S+C21}, where it was assumed that the source position distribution, which is related to the selection function, was known exactly by the observer.

The population model is then described by the set of free parameters introduced in Equations \ref{eq:aterm}, \ref{eq:hterm}, \ref{eq:gterm}, \ref{eq:muh}, and \ref{eq:mug}, plus the Hubble constant $H_0$:
\begin{equation}\label{eq:hyperpars}
\hyperpars \equiv \left\{H_0,\asps,\mu_{\mathrm{DM},0},\beta_{\mathrm{DM}},\xi_{\mathrm{DM}},\sigma_{\mathrm{DM}},\mu_\gamma,\beta_\gamma,\xi_\gamma,\sigma_\gamma\right\}.
\end{equation}
\Tref{tab:inference} provides a brief description of each parameter.

I point out that this model differs from the one used to generate the mock, described in \Sref{sect:sims}, in two key aspects: (1) the true dark matter density profile is not a gNFW model with fixed scale radius; and (2) the distributions in the projected dark matter mass within $5$~kpc and inner dark matter slope do not  strictly follow Equations \ref{eq:hterm} and \ref{eq:gterm}.
These differences make the experiment realistic: in a real-world application, it is unlikely that a simply parameterised model can\textbf{} reproduce the dark matter distribution of a sample of galaxies  exactly. 

\subsection{Inference technique}

I want to calculate the posterior probability distribution of the model parameters, $\hyperpars$, given the data $\data$. Using Bayes' theorem, this is
\begin{equation}\label{eq:bayes}
\pr(\hyperpars|\data) \propto \pr(\hyperpars)\pr(\data|\hyperpars),
\end{equation}
where $\pr(\hyperpars)$ is the prior probability of the parameters and $\pr(\data|\hyperpars)$ the likelihood of observing the data given the model. As measurements carried out on different lenses are independent of each other, the latter is the following product:
\begin{equation}\label{eq:likelihood}
\pr(\data|\hyperpars) = \prod_i \pr(\datai|\hyperpars),
\end{equation}
where $\datai$ is the data relative to lens $i$.

The parameters $\hyperpars$ do not directly predict the data: those depend on the values of $\indpar$ taken by the individual lenses. In order to evaluate each product $\pr(\datai|\hyperpars)$, it is therefore necessary to marginalise over all possible values of $\indpar$:
\begin{equation}\label{eq:fullintegral}
\pr(\datai|\hyperpars) = \int d\indpari \pr(\datai|\indpari,\hyperpars) \pr(\indpari|\hyperpars).
\end{equation}
Formally, this is a six-dimensional integral. A similar calculation to that performed in \citetalias{S+C21} allows one to reduce it to the following:
\begin{equation}\label{eq:2dintegral}
\begin{split}
\pr(\datai|\hyperpars) = & \int d\gammadm \int d\log{\left(\frac{\mfive}{1/\hseventy^2}\right)} \left\lvert\detJ\right\rvert_{\left(\frac{\mstartrueein}{1/\hseventy^2},\betaein\right)}\\
& \pr\left(\rmurobs \left\vert\gammadm,\frac{\mfive}{1/\hseventy^2},\frac{\reff}{1/\hseventy},\frac{\mstartrueein}{1/\hseventy^2},\betaein\right.\right) \\
& \pr\left(\dtobs\left\vert H_0,\gammadm,\frac{\mfive}{1/\hseventy^2},\frac{\reff}{1/\hseventy},\frac{\mstartrueein}{1/\hseventy^2},\betaein\right. \right) \\
& \pr\left(\frac{\mobs}{1/\hseventy^2}\left\vert \frac{\mstartrueein}{1/\hseventy^2},\asps\right. \right) \\
& \pr\left(\left.\frac{\mstartrueein}{1/\hseventy^2},\frac{\reff}{1/\hseventy},\frac{\mfive}{1/\hseventy^2},\gammadm,\betaein\right\vert \hyperpars\right).
\end{split}
\end{equation}
In the integrand function, $\hseventy^2\mstartrueein$, and $\betaein$ are the values of the stellar mass and source position needed to reproduce the two image positions given the dark matter parameters $\mfive$ and $\gammadm$. The factor $\detJ$ is the Jacobian determinant corresponding to the variable change from $(\log{(\hseventy^2\mstartrue)},\beta)$ to the image positions $(\theta_1,\theta_2)$, evaluated at $\hseventy^2\mstartrueein$ and $\betaein$.

I assumed uniform priors on all of the population parameters in \Eref{eq:hyperpars}, with the exception of $\asps$, for which I assumed a uniform prior on its base-ten logarithm.
The bounds on each parameter are listed in \Tref{tab:inference}.
I sampled the posterior probability distribution $\pr(\hyperpars|\data)$ with {\sc emcee} \citep{For++13}, a Python implementation of the \citet{G+W10} affine-invariant sampling method.
At each draw of the parameters $\hyperpars$, I calculated the integrals of \Eref{eq:2dintegral} numerically by computing the integrand function on a grid and doing a spline interpolation and integration over the two dimensions.
I verified that the inference method is accurate within the uncertainties by applying it to mock lens populations generated with the same properties as the model fitted to them.


\section{Results}\label{sect:results}

Before showing the results of the inference, it is useful to make a few general considerations on the degree of precision that we can expect given the dataset.
A single lens in the sample provides a time-delay measurement with $10\%$ precision. This means that, if the lens model parameters were known exactly, this single measurement could be used to obtain an estimate of $H_0$ with the same precision.

The statistical combination of $100$ such lenses would reduce the uncertainty to $1\%$.
This is the highest possible precision attainable in the ideal case with no uncertainties related to the lens model. However, in this experiment, not only does one need to determine the lens model parameters from noisy data, but also these parameters are underconstrained on an individual lens basis: by fitting the model of section \ref{ssec:individ} to a single lens, one obtains a posterior probability that is dominated by the prior (section \ref{ssec:whatbayes} shows this more quantitatively).
This means that a single-lens inference on $H_0$ carries, in addition to the uncertainty on the time delay, an uncertainty related to the lens model parameters, which in turn depends on their prior probability distribution.

When combining the $100$ lenses in the sample with a hierarchical inference formalism, the population distribution of \Eref{eq:popdist} acts as a prior on the individual lens parameters $\indpar$. This distribution is not fixed, but its parameters $\hyperpars$ are inferred from the data at the same time as $H_0$.
In summary, the final uncertainty on $H_0$ will depend on (1) the uncertainties on the time-delay measurements, (2) the uncertainties associated with the lens models of the individual lenses, and (3) the uncertainties on the population distribution parameters $\hyperpars$. 
Each one of these aspects can dominate over the others, depending on the sample size, data quality, and model complexity. As I show in this section, (3) is the dominant source of error on $H_0$ in this experiment.

\subsection{Inference from 100 time-delay lenses}\label{ssec:100alone}

\begin{figure*}
\includegraphics[width=\textwidth]{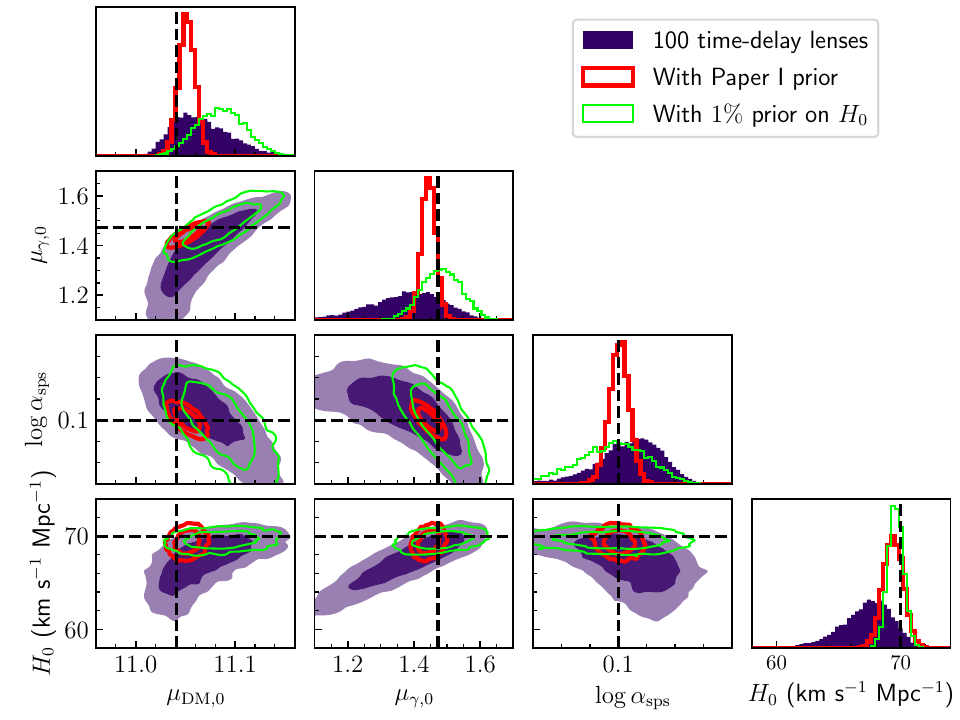}
\caption{
Posterior probability distribution of the four key parameters of the model: the Hubble constant, the stellar population synthesis mismatch parameter, the average $\log{\mfive}$, and the average $\gammadm$. 
Purple filled contours correspond to the fit to the sample of 100 time-delay lenses, with no extra information.
Red lines show the posterior probability obtained by using prior information on the model parameters from the sample of 1000 strong lenses simulated in \citetalias{S+C21}.
Contour levels correspond to 68\% and 95\% enclosed probability regions.
Dashed lines indicate the true values of the parameters, which are defined by fitting the model directly to the distribution of $\log{\mfive}$, $\gammadm$, and $\log{\asps}$ of the mock sample.
\label{fig:inference}
}
\end{figure*}

\Fref{fig:inference} shows the posterior probability distribution of the main parameters of the model: the Hubble constant, the stellar population synthesis mismatch parameter, the average dark matter mass within $5$~kpc, and the average dark matter slope.
The median and 68\% credible bounds of the marginal posterior probability of each parameter are reported in \Tref{tab:inference}.
I defined the true values of the model parameters by fitting the model distribution of section \ref{ssec:popdist} directly to the values of $\mfive$ and $\gammadm$ of the mock sample, which I obtained by fitting a gNFW profile to the projected surface mass density of each lens.
\begin{table*}
\caption{Inference on the model parameters.
\label{tab:inference}
}
\begin{tabular}{lcccccl}
\hline
\hline
Parameter & Truth & Prior & 100 time-delay & With prior   & With $H_0$ & Description \\
          &       &       & lenses         & from \citetalias{S+C21} & prior      & \\
\hline
$H_0$ & $70$ & $U(50,90)$ & $67.5_{-2.2}^{+1.7}$ & $69.4_{-0.8}^{+0.8}$ & $69.6_{-0.6}^{+0.6}$ & Hubble constant, in km s$^{-1}$ Mpc$^{-1}$ \\
$\mu_{\mathrm{DM}, 0}$ & $11.04$ & $U(10,12)$ & $11.06_{-0.02}^{+0.03}$ & $11.052_{-0.008}^{+0.008}$ & $11.09_{-0.02}^{+0.02}$ & Mean $\log{\mfive}$ at $\log{\msps}=11.4$ and \\
 & & & & & & average size \\
$\beta_{\mathrm{DM}}$ & $0.57$ & $U(0,3)$ & $0.49_{-0.06}^{+0.06}$ & $0.56_{-0.02}^{+0.02}$ & $0.53_{-0.05}^{+0.06}$ & Dependence of $\log{\mfive}$ on $\msps$ \\
$\xi_{\mathrm{DM}}$ & $-0.15$ & $U(-1,1)$ & $-0.20_{-0.12}^{+0.10}$ & $-0.13_{-0.04}^{+0.04}$ & $-0.21_{-0.11}^{+0.09}$ & Dependence of $\log{\mfive}$ on galaxy size \\
$\sigma_{\mathrm{DM}}$ & $0.05$ & $U(0.0,0.5)$ & $0.032_{-0.009}^{+0.014}$ & $0.058_{-0.008}^{+0.008}$ & $0.035_{-0.011}^{+0.017}$ & Intrinsic scatter in $\log{\mfive}$ \\
$\mu_{\gamma,0}$ & $1.47$ & $U(0,1)$ & $1.38_{-0.12}^{+0.10}$ & $1.44_{-0.02}^{+0.02}$ & $1.48_{-0.06}^{+0.06}$ & Mean $\gammadm$ at $\log{\msps}=11.4$ and \\
 & & & & & & average size \\
$\beta_{\gamma}$ & $-0.39$ & $U(-1,1)$ & $-0.37_{-0.10}^{+0.10}$ & $-0.27_{-0.04}^{+0.04}$ & $-0.42_{-0.10}^{+0.10}$ & Dependence of $\gammadm$ on $\log{\msps}$ \\
$\xi_{\gamma}$ & $-0.37$ & $U(-1,1)$ & $-0.4_{-0.2}^{+0.2}$ & $-0.28_{-0.08}^{+0.08}$ & $-0.47_{-0.19}^{+0.18}$ & Dependence of $\gammadm$ on galaxy size \\
$\sigma_{\gamma}$ & $0.06$ & $U(0.0,0.5)$ & $0.08_{-0.04}^{+0.05}$ & $0.059_{-0.017}^{+0.016}$ & $0.07_{-0.03}^{+0.03}$ & Intrinsic scatter in $\gammadm$ \\
$\log{\alpha_{\mathrm{sps}}}$ & $0.10$ & $U(0.00,0.25)$ & $0.11_{-0.02}^{+0.02}$ & $0.101_{-0.007}^{+0.007}$ & $0.10_{-0.03}^{+0.02}$ & Log of the stellar population synthesis \\
 & & & & & & mismatch parameter \\\end{tabular}
\tablefoot{
Column (2): true values of the population parameters. For the parameters relative to the dark matter component, these are defined by fitting the model directly to the distribution of $\mfive$ and $\gammadm$. 
Column (3): priors on the parameters.
Columns (4)-(6): median, 16th, and 84th percentile of the marginal posterior probability distribution of each parameter obtained using the sample of 100 time-delay lenses on its own, in combination with a prior on the lens structural parameters from the analysis of \citetalias{S+C21}, and in combination with a prior on $H_0$ with 1\% precision.
}
\end{table*}

Most of the lens structure parameters are recovered within $1\sigma$. However, the true value of the Hubble constant is outside of the inferred 68\% credible region: the measured value is $H_0=67.5_{-2.2}^{+1.7}$~km~s$^{-1}$Mpc$^{-1}$.
This corresponds to a statistical uncertainty of $\sim3\%$, with a systematic bias of $\sim -4\%$.
I verified that the bias is significant by repeating the analysis on new samples of $100$ lenses generated with the same procedure as outlined in \Sref{sect:sims} but with different noise realisations.
The origin of this bias must be searched for in the differences between the truth and the model used to fit the sample: these differences are the dark matter density profile, the population distribution of the dark matter parameters, and the assumed source position distribution.

As we can see from \Fref{fig:inference}, the uncertainty on $H_0$ is due in large part to a strong degeneracy with the parameters $\asps$, $\mu_{\mathrm{DM,0}}$, and $\mu_{\gamma,0}$.
In the following section, I investigate whether the precision and the accuracy on $H_0$ can be improved by making use of prior information from a separate lens sample.

\subsection{Inference with a prior from a sample of 1000 strong lenses}\label{ssec:paper1prior}

\citetalias{S+C21} simulated a statistical strong-lensing measurement on a sample of 1000 lenses. This showed how, with such a sample, it is possible to measure the average properties of the lens population with much higher precision than that obtained with the sample of 100 time-delay lenses considered in this work so far.
For example, the average stellar population synthesis mismatch parameter $\asps$ was recovered with a precision of $\sim0.01$~dex compared to the $\sim0.03$~dex precision obtained in section \ref{ssec:100alone}.
In this section, I investigate a scenario in which the information from a such a sample of 1000 lenses is used as a prior on the parameters describing the lens population in combination with the 100 time-delay lenses simulated in this work.

I proceeded as follow. I repeated the analysis of section \ref{ssec:100alone} using the posterior probability distribution from \citetalias{S+C21} as a prior on the model parameters $\hyperpars$ \citepalias[with the exception of $H_0$, which is unconstrained by the analysis of][]{S+C21}.
There is an important caveat with this approach: I implicitly assumed that the two lens samples are drawn from the same population of galaxies.
To simplify the calculations, I approximated this prior probability distribution as a multivariate Gaussian, with mean and covariance matrix equal to those of the sample obtained from the Markov Chain Monte Carlo of \citetalias{S+C21}. 
The resulting posterior probability distribution in the key parameters is shown as solid red contours in \Fref{fig:inference}, while the median and 68\% credible region of the marginal posterior probability of all parameters is reported in \Tref{tab:inference}.

Remarkably, $H_0$ is now recovered with a precision slightly above $1\%$.
As pointed out in \Sref{sect:results}, $1\%$ is the maximum attainable precision on $H_0$ in the case of perfect knowledge of the lens model parameters, as this is the amplitude of the uncertainty associated with the time-delay measurements alone.
This result tells us that, once the parameters describing the lens population are known with sufficient precision, the main remaining source of uncertainty on $H_0$ is that related to the measurements of $\Delta t_{2,1}$.
This is a key advantage of using a Bayesian hierarchical approach, which is further illustrated in section \ref{ssec:whatbayes}.

\subsection{Inference with a narrow prior on $H_0$}

There are many different ways of measuring the Hubble constant. Therefore, there is the possibility that $H_0$ will be determined with high precision with an experiment different from time-delay lensing.
In that case, measurements of time delays could be used to constrain the properties of the lens population.
I investigated this scenario by repeating the analysis of the 100 time-delay lenses of section \ref{ssec:100alone}, with the addition of a prior on $H_0$ with $1\%$ precision.
The resulting posterior probability distribution is shown in green in \Fref{fig:inference}.

The main improvement brought by the prior knowledge of $H_0$ is in the precision on the inferred average dark matter slope parameter, $\mu_{\gamma,0}$: its uncertainty becomes $0.06$, which is much smaller than the value of $0.12$ obtained in the fiducial analysis.
The other parameters are relatively unchanged: the stellar population synthesis mismatch parameter $\asps$, for instance, is determined with a $0.02-0.03$~dex precision, independently of whether or not an $H_0$ prior is applied.

\subsection{What Bayesian hierarchical inference does}\label{ssec:whatbayes}

Early statistical analyses of time-delay lenses \citep[that is, prior to the][ study]{Bir++20} consisted in obtaining estimates of the time-delay distance $\ddt$ separately for each lens system by marginalising each posterior probability distribution over the parameters describing the lens mass and then combining the resulting marginal posterior probabilities to infer $H_0$ \citep[see e.g.][]{Bon++17}.
Here, I investigate how this approach compares to the Bayesian hierarchical inference method in terms of precision and accuracy.

Firstly, I calculated for each lens the marginal posterior probability distribution in $H_0$ given the data, $\pr(H_0|\datai)$. Formally, this is given by
\begin{equation}\label{eq:onelensh0}
\pr(H_0|\datai) \propto \pr(H_0)\int d\indpari \pr(\datai|\indpari,H_0)\pr(\indpari).
\end{equation}
The term $\pr(\indpari)$ is the prior on the individual lens parameters. In a Bayesian hierarchical approach, this is a function of population-level parameters $\hyperpars$, while in this context the prior is fixed. I assumed a flat prior on the following parameters:
\begin{align}
\log{\msps} \sim & U(10,13) \nonumber \\
\log{\asps} \sim & U(0.0, 0.2) \label{eq:flatpriors} \\
\log{\mfive} \sim & U(10,12) \nonumber \\
\gammadm \sim & U(0.2, 1.8), \nonumber
\end{align}
and the same prior as \Eref{eq:betadist} on the source position $\beta$.

\begin{figure}
\includegraphics[width=\columnwidth]{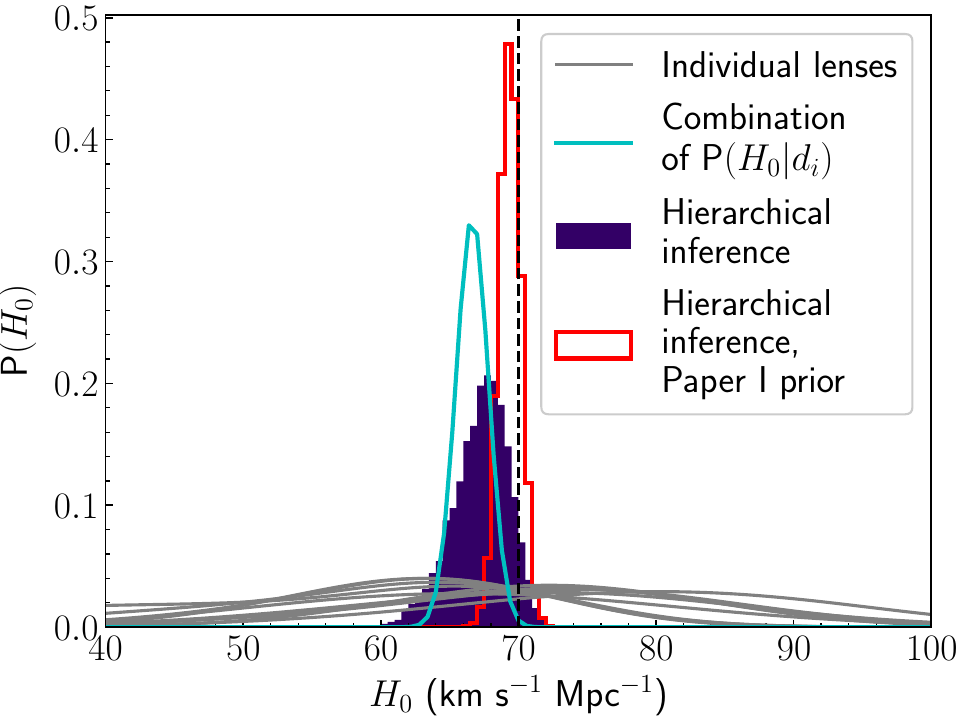}
\caption{
Marginal posterior probability on $H_0$.
Grey curves: Individual lens $\pr(H_0|\datai)$ obtained from \Eref{eq:onelensh0} assuming flat priors on the lens model parameters, for ten time-delay lenses.
Cyan curve: Statistical combination of the marginal posterior probabilities from the 100 time-delay lenses obtained following \Eref{eq:flatpriors}.
Filled purple histogram: Hierarchical inference from section \ref{ssec:100alone}
Red histogram: Hierarchical inference with a prior from \citetalias{S+C21}, from section \ref{ssec:paper1prior}.
The vertical dashed line marks the true value of $H_0$ used to simulate the data.
\label{fig:marginal}
}
\end{figure}

\Fref{fig:marginal} shows $\pr(H_0|\datai)$ for ten lenses of the sample (grey curves). The $1\sigma$ uncertainty on $H_0$ is typically around $15\%-20\%$. 
If I consider this to be the result of the quadrature sum between the error on the time delay, which is $10\%$, and that on the individual lens parameters, it follows that the latter is the dominant source of uncertainty. This is not surprising, because the lens model is under-constrained.

I then considered the joint inference on $H_0$, which was obtained by combining the marginal posterior probabilities of the 100 lenses of the sample. This is defined as
\begin{equation}\label{eq:simplecomb}
\pr(H_0|\data) \propto \pr(H_0) \prod_i \int d\indpari \pr(\datai|\indpari,H_0)\pr(\indpari).
\end{equation}
The resulting posterior probability distribution is shown as a cyan curve in \Fref{fig:marginal}.
The value of the Hubble constant inferred in this way is $H_0 = 66.6\pm1.2\,\rm{km}\,\rm{s}^{-1}\,\rm{Mpc}^{-1}$. The relative uncertainty on $H_0$ is $1.7\%$, a factor $1/\sqrt{N}$ smaller than the individual lens measurements, which is the standard result when $N$ independent measurements of a given quantity are combined.
However, the inference is highly biased.

This is a good example of a result driven by the prior.
The choice of prior on the dark matter slope made with \Eref{eq:flatpriors} is particularly detrimental in this case: $\pr(\gammadm)$ is centred at $\gammadm=1$, corresponding to an NFW profile, and gives equal probability to values of $\gammadm$ above or below it. While this appears to be a reasonable choice, the true dark matter slope of the lenses is on average much steeper than that of an NFW profile. As a result, the inferred value of $H_0$ is biased low. I verified that, repeating this analysis with a higher lower bound on $\gammadm$, the inference on $H_0$ gets closer to the truth.

With the Bayesian hierarchical approach, the inference on $H_0$ is more uncertain, but also more accurate: the true value of $H_0$ is recovered within $2\sigma$.
This is because the hierarchical model is designed in such a way that one can infer the probability distribution of the individual lens parameters, $\pr(\indpar|\hyperpars)$, along with $H_0$, instead of assuming a fixed prior on them. As a result, the uncertainty on quantities such as the average dark matter slope (parameter $\mu_{\gamma,0}$) propagates over the inference on $H_0$ and ends up being the dominant source of error.
The data from the  100 simulated lenses does not allow us to distinguish between a scenario in which (a) $H_0$ is close to the true value and the average $\gammadm$ is $1.5$, and one where (b) $H_0\approx65\,\rm{km}\,\rm{s}^{-1}\,\rm{Mpc}^{-1}$ and $\mu_{\gammadm,0}\approx1.2$. This uncertainty is properly taken into account by the hierarchical model.
In other terms, while the simple statistical combination of \Eref{eq:simplecomb} treats each lens independently from the others, the hierarchical model takes into account correlated errors due to our ignorance of the average properties of galaxies.

Finally, the Bayesian hierarchical method can also provide higher precision than a traditional approach when prior information on the lens population parameters is available. 
This is illustrated by the example of section \ref{ssec:paper1prior}, which makes use of the prior from the sample of \citetalias{S+C21}.
The resulting marginal posterior on $H_0$, shown as a red curve in \Fref{fig:marginal}, has a $1\sigma$ uncertainty of $1.3\%$, which is significantly smaller than the $1.7\%$ uncertainty obtained when assuming flat priors on the individual lens parameters.
This is because, if the population-level parameters $\hyperpars$ are known with sufficient precision, lens models corresponding to the tails of the $\pr(\indpar|\hyperpars)$ distribution are suppressed, resulting in a more precise measurement.


\section{Discussion and summary}\label{sect:discuss}

The experiments carried out in this work were realised under a series of simplifying assumptions.
In order to apply the statistical inference method developed here to a real sample of lenses, there are several challenges to be overcome. 
Many of these challenges apply in the same way to the analysis of \citetalias{S+C21}. These are: generalising the lens model to the non-axisymmetric case and modelling the full surface brightness distribution of the lensed source and computing the individual lens parameter marginalisation integrals of \Eref{eq:fullintegral} in a computationally efficient yet accurate way.
Dropping the axisymmetric lens assumption can be particularly challenging, as it requires increasing the dimensionality of the problem.
I refer to section 6.5 of \citetalias{S+C21} for a thorough discussion of these points.

One important aspect in time-delay lensing studies that I have left out in this work is the line-of-sight structure. The effect of mass perturbers along the line of sight can usually be described with a constant sheet of mass. As such, the line-of-sight structure has the same impact on the time delay as the mass-sheet transformation described in \ref{ssec:masssheet}: if not taken into account, it will bias the inference on $H_0$.
In state-of-the-art time-delay lensing studies, the effect of line-of-sight perturbers is modelled on the basis of measurements of the environment of each lens \citep[see e.g.][ for details]{Rus++17}.
In principle, the modelling of the line of sight could be incorporated into the same hierarchical model describing the lens galaxy population, although that would complicate the analysis. 

The first result of this study is the finding that, with a sample of 100 doubly lensed quasars, each with a $10\%$ precision measurement of the time delay and with high-resolution imaging data, the Hubble constant can be measured with a precision of about $3\%$.
The main source of uncertainty is the knowledge of the distribution of the lens structural parameters: the stellar mass-to-light ratio and the dark matter density profile.
However, the experiment also showed that the inference can suffer a bias of comparable amplitude.
The origin of this bias lies most likely in the fact that the model that I used for the inference, although relatively flexible, is not a perfect description of the reality assumed when creating the simulated sample of lenses.
While I cannot exclude that a higher degree of accuracy could be reached with alternative models, it is clear that in order to obtain a $1\%$ measurement of $H_0$, additional information is needed, either from a much larger sample of time-delay lenses, from external datasets, or from predictions from hydrodynamical simulations \citep[such as with the method proposed by][]{Har20}.

The LSST should be able to provide the data necessary to measure time delays for about 400 strongly lensed quasars \citep{Lia++15}, with the actual number depending on the observing strategy and the required precision on these measurements \citep[see also section 5.2 of][]{Loc++21}.
Such a sample would allow to achieve a substantially higher precision on the inferred lens population parameters, and consequently on the Hubble constant, compared to the 100-lens scenario examined here. However, a quantitative forecast of the constraining power of LSST time-delay lenses is beyond the goals of this work.

The second result is that, when prior information on the lens structure distribution from a large external lens sample is combined to a set of 100 time-delay lenses, it is possible to reach a precision and accuracy of about $1\%$. 
This result also showed that, when the lens population parameters are known very well, uncertainties associated with the modelling of individual lenses are greatly reduced in virtue of the knowledge of their probability distribution.

Combining time-delay lenses with larger samples of regular strong lenses (i.e. with no time delays) is the strategy currently pursued by the TDCOSMO collaboration \citep{Bir++20,B+T21}.
\citet{B+T21} forecast that a joint sample of 40 time-delay lenses and 200 regular lenses can lead to a $1.5\%$ precision on $H_0$. The main difference between their work and this one is that \citet{B+T21} relied on stellar kinematics measurements to further constrain the lens model parameters. While a larger sample of lenses is needed to achieve the same precision with the approach of the present paper, the advantage of not relying on stellar kinematics is that the inference is immune to possible systematic effects associated with the stellar dynamical modelling step.
In order to have a good handle on systematic errors, it is therefore worth pursuing both of these strategies.

An implicit assumption made in both this and the \citet{B+T21} study is that the samples of time-delay and non-time-delay lenses are drawn from a population of galaxies with the same properties.
However, differences in the selection criteria can mean that the two samples probe different subsets of the general galaxy population, with potentially different underlying distributions in the lens structural parameters.
When applying this method in practice, it is therefore important to either make sure that these differences do not introduce significant biases, or to explicitly model the selection effects relative to both lens samples.
\citet[][, Paper III]{Son21b} provides a framework for taking lens selection effects into account in a Bayesian hierarchical inference.

Finally, I show that if a $1\%$ measurement of $H_0$ is available from a separate experiment, then a sample of 100 time-delay lenses can be used to constrain the properties of the mass structure of the lens population. The main effect of the prior information on $H_0$ is that of improving the determination of the average dark matter density slope.
Stellar mass measurements can be calibrated with a $0.03$~dex precision with such a sample, regardless of prior knowledge on $H_0$.

The strong lensing data simulated in this study consisted of image positions, magnification ratios, and time delays. 
However, if the background source is a standardisable candle, such as a type 
Ia supernova, it is possible to obtain information on the absolute magnification, which can break the mass-sheet degeneracy. A scenario in which such magnification measurements are available was explored by \citet{BDS21} with promising results.

In summary, time-delay lenses are very powerful probes of both cosmology and galaxy structure. The LSST will provide data that will enable time-delay measurements for hundreds of lenses. This work presents a strategy to exploit these data with minimal use of external (non-lensing) information.


\begin{acknowledgements}

I thank Phil Marshall, Marius Cautun, Simon Birrer, Timo Anguita and Frederic Courbin for useful discussions and suggestions.

\end{acknowledgements}

\bibliographystyle{aa}
\bibliography{references}

\appendix
\section{Mass-sheet transformations of the lens model}\label{sect:appendixa}

When modelling a time-delay lens for the purpose of measuring $H_0$, it is important to ensure that the assumed lens model family does not break the mass-sheet degeneracy. 
This means verifying that, when applying a mass-sheet transformation of the kind of \Eref{eq:mst} to the lens model, the transformed model still belongs to the original model family.
Strictly speaking, this is not the case for the model adopted in this paper (or any physically motivated model). 
For instance, the dimensionless surface mass density changes as follows under a mass-sheet transformation:
\begin{equation}
\kappa \rightarrow \kappa' = \lambda\kappa + 1 - \lambda.
\end{equation}
When $\lambda > 1$, $\kappa'$ becomes negative at large radii, where $\kappa$ approaches zero.
A negative surface mass density cannot be produced by the lens model of section \ref{ssec:individ}, for any combination of parameters.

However, for the purpose of ensuring accuracy in the inference of $H_0$, it is sufficient to show that the mass-sheet degeneracy is not broken over the region constrained by the lensing data.
The relevant quantities are then the local derivatives of the lens potential at the Einstein radius.
In \citet{Son18} I showed that, for small displacements around the Einstein radius, the time delay is proportional to the product of the first and second derivative of the potential, $\psi'\psi''$, where $\psi'$ is equal to the Einstein radius. The image positions and the radial magnification ratio, instead, only constrain the first derivative $\psi'$ and the following combination of derivatives:
\begin{equation}\label{eq:invariant}
\frac{\psi'''}{1 - \psi''}.
\end{equation}
A mass-sheet transformation has the following effect on the first three derivatives of the potential:
\begin{align}\label{eq:son18}
\psi' \rightarrow & \psi' \nonumber \\
\psi'' \rightarrow & \lambda\psi'' + 1 - \lambda \\
\psi''' \rightarrow & \lambda\psi''' \nonumber,
\end{align}
from which it follows that the quantity in \Eref{eq:invariant} does not change under a mass-sheet transformation (both the numerator and the denominator scale with $\lambda$).
If the lens model allows for variations in the potential derivatives of the kind of \Eref{eq:son18}, then it does not break the mass-sheet degeneracy.

To check whether this condition is satisfied, I considered the following example.
I generated a lens using the model of section \ref{ssec:individ}, with a stellar mass of $\log{\mstar}=11.5$, a half-light radius $\reff=7\,{\rm kpc}$, a dark matter mass $\log{\mfive}=11.0,$ and dark matter slope $\gammadm=1.5$. The Einstein radius of this lens (for the same lens and source redshift used in the main experiment) is $\psi'=1.15''$.
I then varied the three lens model parameters while keeping the Einstein radius fixed, examined the corresponding variation in the lens potential derivatives, and compared this with that produced by a mass-sheet transformation.

\begin{figure}
\includegraphics[width=\columnwidth]{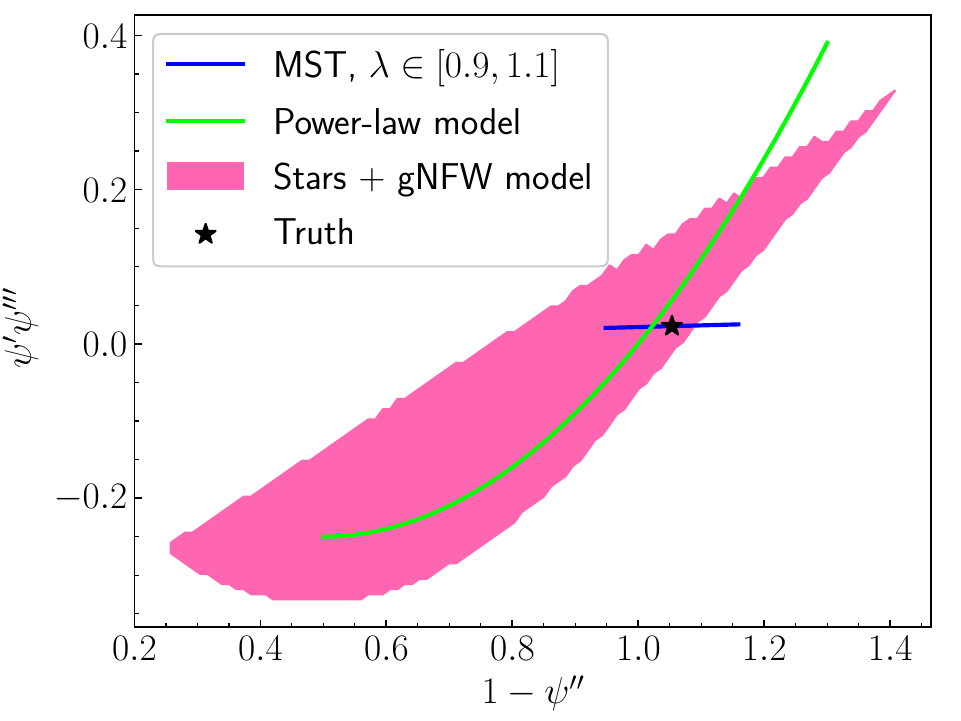}
\caption{
Degrees of freedom of the model and the mass-sheet transformation.
Values of the product between the third and first derivative of the lens potential, evaluated at the Einstein radius, $\psi'\psi'''$, as a function of $1-\psi''$, for various lenses with fixed Einstein radius.
The star marks the values of a lens generated with the model of section \ref{ssec:individ}, assuming $\log{\mstar}=11.5$, $\reff=7\,{\rm kpc}$, $\log{\mfive}=11.0$, $\gammadm=1.5$.
The pink region covers the range of values spanned by the model, obtained by varying the three model parameters while keeping the Einstein radius fixed.
The blue line indicates the range of values obtained by applying the mass-sheet transformation of \Eref{eq:mst} to the original lens, with values $\lambda\in[0.9,1.1]$.
The green line corresponds to a power-law lens model.
\label{fig:psipsi}
}
\end{figure}

\Fref{fig:psipsi} shows the distribution in the (dimensionless) product $\psi'\psi'''$ as a function of $1-\psi''$ covered by the lens model (pink region).
The true values of these quantities (those corresponding to the lens generated initially) are marked by a star. 
A mass-sheet transformation changes both $\psi'\psi'''$ and $1 - \psi''$ by a factor $\lambda$. The purple line shows such a transformation over the range $\lambda\in[0.9,1.1]$.
The transformed lens falls in the region covered by the lens model, with the exception of values of $\lambda > 1.03$.
Large positive values of $\lambda$ can anyway be excluded, because they correspond to unphysical models, with negative convergence at large radii.
I therefore conclude that the lens model used in the analysis does not break the mass-sheet degeneracy for physically acceptable values of $\lambda$.

\end{document}